\documentclass[12pt]{iopart}
\usepackage{iopams}
\usepackage[utf8]{inputenc}
\usepackage[english]{babel}
\usepackage{graphicx}

\usepackage[numbers, sort&compress]{natbib}
\usepackage{ulem}
\usepackage{blindtext}
\expandafter\let\csname equation*\endcsname\relax

\expandafter\let\csname endequation*\endcsname\relax
\usepackage{import}
\usepackage{amsmath}
\usepackage{graphicx}
\usepackage[usenames,dvipsnames]{xcolor}
\usepackage{caption}
\usepackage[braket, qm]{qcircuit}
\usepackage{tikz}
\usepackage{float}
\usepackage{wrapfig}
\usepackage{xargs}
\usepackage{textcase}
\usepackage{url}
\usepackage{bm}
\usepackage[colorlinks]{hyperref}
\hypersetup{
    colorlinks=true,
    citecolor=blue,
    linkcolor=red,
    filecolor=red,
    urlcolor=blue,
}
\usepackage[margin=1in]{geometry}
\usepackage{subcaption}
\usepackage{lineno}
\def\beq{\begin{equation}}
\def\enq{\end{equation}}
\begin{document}
\title[Indexed improvements for real-time Trotter evolution]{Indexed improvements for real-time Trotter evolution of a $(1+1)$ field theory using NISQ quantum computers}

\author{Erik Gustafson (erik-j-gustafson@uiowa.edu)}
\address{Department of Physics and Astronomy, The University of Iowa, Iowa City, IA 52242, USA}
\author{Patrick Dreher}
\address{Department of Computer Science, North Carolina State University, Raleigh, NC 27695, USA}
\address{Department of Physics, North Carolina State University, Raleigh, NC 27695, USA}
\author{Zheyue Hang}
\address{Department of Physics and Astronomy, The University of Iowa, Iowa City, IA 52242, USA}
\author{Yannick Meurice}
\address{Department of Physics and Astronomy, The University of Iowa, Iowa City, IA 52242, USA}

\begin{abstract}
Today's quantum computers offer the possibility of performing real-time calculations for quantum field theory scattering processes motivated by high energy physics. In order to follow the successful roadmap which has been  established for the calculation of static properties at Euclidean time, it is crucial to develop new algorithmic methods to deal with the limitations of current Noisy Intermediate-Scale Quantum (NISQ) devices and to establish quantitative measures of the progress made with different devices. In this paper, we report recent progress in these directions. We show that nonlinear aspects of the Trotter errors allow us to take much larger step then suggested by low-order analysis. This is crucial to reach physically relevant time scales with today's NISQ technology. We propose to use an index averaging absolute values of the difference between the accurately calculated Trotter evolution of site occupations and their actual measurements on NISQ machines (G index) as a measure to compare results that have been obtained from different hardware platforms. Using the transverse Ising model in one spatial dimension with four sites we apply this metric across several hardware platforms.  We study the results including readout mitigation and Richardson extrapolations and show that the mitigated measurements are very effective based on the analysis of the Trotter step size modifications.  We discuss how this advance in the Trotter step size procedures can improve quantum computing physics scattering results and how this technical advance can be applied to other machines and noise mitigation methods.

\end{abstract}
\maketitle

\section{Introduction}

Euclidean time formulations of field theory models of strongly correlated systems using Monte Carlo sampling have  been very successful for determining these model's static properties. An important example is lattice quantum chromodynamics (QCD), a non-perturbative formulation of the theory of strong interactions among quarks and gluons.  However, Monte Carlo sampling is ineffective to deal with {\it real-time} evolution in field theories because of the sign problem caused by oscillating phases.  In addition, using the full Hilbert space of the 
Hamiltonian formulation with a classical computer requires resources that grow exponentially with the size of the system. 

Unlike a classical computer, a quantum computer is based on a fundamentally different computational paradigm. It can handle large Hilbert spaces with a cost growing polynomially with the size of the system \cite{Lloyd1073} and  has no sign problem. Using quantum computers to perform ab-initio calculations of the real-time evolution of quarks and gluons based on lattice QCD  is a strategic,  long-term, goal for the high-energy and nuclear physics community.  Successfully implementing this strategy can have potentially high impact on the interpretation of particle collider experiments, nuclear spectra and compact object astrophysics. In addition, this new approach could be used for finite-density calculations which also suffer from a sign problem.  However, quantum computing technology is in its early stages of development and today's machines are best described as Noisy Intermediate-Scale Quantum (NISQ) computing hardware platforms.  At the present time these limited capability machines can only handle the real-time evolution of simple quantum field theories such as the quantum spin or gauge Ising models in small spatial volumes. 

The lattice gauge theory community has successfully established a roadmap \cite{RevModPhys.51.659} starting with Ising models and culminating with realistic  calculations of  masses and form factors using lattice quantum chromodynamics.  We plan to follow the same roadmap, a sequence of models of increasing complexity sometimes called the ``Kogut ladder", but now for the real-time evolution in quantum field theory. Progress towards this goal  requires algorithmic and technological improvements, error mitigation as well as quantitative methods to assess the progress made. In this paper we report progress in these directions. We discuss nonlinear aspects of the Trotter error that allow us to use take larger steps than suggested by low order bounds
and reach larger time scale. We introduce an index that measures the closeness of NISQ data to numerical 
results that can be computed accurately for sufficiently small systems. We apply this index to compare 
devices and error mitigation methods. Overall, we demonstrate that good progress is being made and we expect significant improvements in the near future. 

The Trotter approximation plays an important part in the proof \cite{Lloyd1073} that for local interactions, quantum computers can handle the real-time evolution for systems with many quantum degrees of freedom. 
The errors can be controlled by taking sufficiently small steps but at the cost of needing more steps to 
reach physically relevant time scales, which is problematic with NISQ devices. 
The paper presents an advance in utilizing the Trotterization selecting the time step $(\delta t)$ $10-20$ times larger than would be suggested by bounds of order $\delta t^2$ and  $\delta t ^3$, in order to optimize the time evolution of these systems on today's NISQ hardware platforms. This was considered in Ref.  \cite{GustafsonIsing} and then documented more systematically \cite{ybook,treview}. Considering these larger Trotter steps involves identifying non-linear aspects such as resonances in the fidelity that may appear 
and studying their impacts on the calculations. We also study the impacts that noise has on these calculations when implemented on hardware platforms with different qubit layout topologies.  For this purpose we introduce an index that allows us to discriminate among mitigation methods and devices. Having a clear metric is crucial to measure the progress made.

One important limitation in today's superconducting transmon devices is that their qubits can only maintain quantum computing coherence on the order of tens to hundreds of microseconds. In effect, the circuit depth available to program these machines is seriously constrained \cite{gilliam2020canonical, pino2020demonstration} and any algorithm and quantum circuit design must factor this limitation into its implementation.  Central to this task is confronting the issues of how to mitigate the noise on these platforms in order to improve the performance and extend the capability of these quantum computations.

To study the adverse impacts of noise on quantum computing calculations, it is important to identify specific processes that are both sensitive to noise on an individual system and that can also be used to compare devices and measure the improvements in successive generations of quantum computing hardware platforms.  The transverse Ising model (TIM) is an excellent candidate for measuring these improvements by modelling real-time field theory scattering processes on these NISQ machines.  It is a local field theory (nearest neighbor interactions) with connections to both condensed matter physics and quantum field theories \cite{RevModPhys.51.659}. The shallow-depth circuits from this model can be easily coded and implemented onto today's quantum computers. They provide a stable production model for studying the real-time evolution of a physical observable.

Extensive work with these type of circuits has already been done simulating \cite{Johanning_2009,Smith_2019,Lamm:2018siq,Tan_2021}, developing algorithmic tools \cite{PhysRevLett.119.180509, somma2016quantum,PhysRevX.7.021050,PhysRevX.8.031027}, and testing them \cite{Kandala_2019,Dumitrescu:2018njn} for this model.  Extensions to more complicated models such as general scalar and fermionic field theories \cite{Jordan:2011ci, Jordan:2011ne, Moosavian:2017tkv,Klco:2018zqz, Macridin:2018oli,Yeter-Aydeniz:2018mix,Roggero:2018hrn},  and gauge-matter theories \cite{Roggero_2020,Holland_2020,Martinez:2016aa,Klco:2018kyo,Raychowdhury:2018osk,Stryker:2018efp,Hackett:2018cel,Muschik:2016tws,Kokail:2018eiw} are also currently being studied. At the present time there are several groups examining various behaviors of the transverse Ising model on quantum computers
\cite{GustafsonIsing,Lamm:2018siq,PhysRevA.79.062314,CerveraLierta2018exactisingmodel}.  Results from these quantum simulators offer a direction for mapping these problems onto today's quantum computing hardware platforms. 
It is not difficult to extend this work to other field theoretical models such as the Thirring and Schwinger models.  

In Sec. \ref{secTheory} we examine the one-dimensional transverse Ising model with four sites and open boundary conditions (OBC).  In Sec. \ref{subsec:trotter}, we discuss the one step Trotter error 
for rather large $\delta t$. We then propose a specific metric and use it to compare three IBM machines
and several methods of error mitigation (readout calibration and Richardson
extrapolation) for the quantum Ising model in one spatial dimension.  More specifically, we have designed a new benchmarking metric, $G_{\epsilon}$ ($G$-index) that improves characterizations of this Trotterization method.  This metric explicitly calculates the difference between the measured site occupations and the exact value, for all site occupations in the case where the measured value is above a certain threshold  $\epsilon$.  This metric helps to gauge how accurately a quantum system, which can be trivially implemented on a quantum computer,\footnote{We define ``trivially implemented" as meaning no unnecessary swap gates are needed to simulate the model.} can be simulated using current quantum computers across different days.\footnote{In \ref{sec:machinespecs} we examine the measurements from different hardware generations and show the consistency across these platforms. Identifying how consistently a NISQ computer performs day to day informs us as to how far in time we can evolve a quantum system \cite{Johanning_2009}.} 

Based on the observed data from these machines in Sec. \ref{secErrorMitigation}, we analyze how this advance in utilizing the Trotterization procedure provides an effective measure on the number of Trotter steps or the depth of a circuit that can be used.  We also discuss how the readout errors can affect the simulation results and how they can be rectified. We use Richardson extrapolation schemes as carried out in \cite{GustafsonIsing,Klco:2018kyo,PhysRevX.8.031027,PhysRevX.7.021050}, to gauge how much noise we can remove via post processing of the data.

In Sec. \ref{secAlgorithmicMitigation}, we examine methods of algorithmic mitigation, where we attempt to reduce the error coming from the Trotter approximation by extrapolation methods. We discuss the non-linear aspects of the evolution operator which creates ``resonances" in the fidelity as a function of the Trotter step $\delta{t}$ and explain that it can lead to problematic extrapolations. Sec. \ref{secConclusion} summarizes our conclusions from this work. We also discuss future applications of the methods described in our article.

\section{Methodology for real-time calculations and benchmarking}
\label{secTheory}
\subsection{Ising Hamiltonian}
The formulation of the transverse Ising model that we examine uses four sites and has OBC. This model can be explicitly written as:
\begin{equation}
\label{eqHamiltonian}
\hat{H} = -J \sum_{i = 1}^3 \hat{\sigma}^x_i \hat{\sigma}^x_{i+1} - h_T \sum_i^4 \hat{\sigma}^z_i,
\end{equation}
where $J$ is the nearest neighbor coupling (hopping) and $h_T$ is the on-site energy. Following Ref.  ~\cite{GustafsonIsing}, we chose $J = 0.02$ and $h_T = 1.0$ because of the
simple connection to single particle quantum mechanics. This is why we use the ``particle basis" where the dominant on-site energy is diagonal. The relevant time scale is discussed in Section~\ref{subsec:trotter}.
We used OBC because Almaden and Boeblingen do not allow a four site Ising model with periodic boundary conditions to be trivially implemented on the quantum hardware (See Fig. \ref{fig:almadenlayout} for the layout of Almaden and Boeblingen and Fig. \ref{fig:melbournelayout} for the layout of Melbourne in \ref{sec:machinespecs}).

\subsection{The Trotter Approximation}
\label{subsec:trotter}

The system can be evolved in time using the complex exponential of the Hamiltonian:
\begin{equation}
\label{eqtimeevolveexact}
\hat{U}(t) = e^{- i t \hat{H}}.
\end{equation}
Following Refs. \cite{Lloyd1073,GustafsonIsing},
the Trotter approximation
is applied to the evolution operator
with the explicit form:
\begin{equation}
\label{eqsuzuki}
\hat{U}(t;N) = \Big(\hat{U}_1(t / N; h_t) \hat{U}_2(t / N; J)\Big)^N + \mathcal{O}(t^2 / N)
\end{equation}
where $N$ is the number of Trotter steps to be implemented, $(\delta t)$ is the Trotter step size
\begin{equation}
\label{eqtfieldevo}
\hat{U}_1(\delta t; h_t) = e^{-i h_T \delta t \sum_{i = 1}^{4} \hat{\sigma}^z_i },
\end{equation}
and
\begin{equation}
\label{eqhoppingevo}
\hat{U}_2(\delta t; J) = e^{-i J \delta t \sum_{i = 1}^{3} \hat{\sigma}^x_i \hat{\sigma}^x_{i+1}}.
\end{equation}
The operators defined in Eqs. \ref{eqtfieldevo}  and \ref{eqhoppingevo} can be expressed as a combination of the following two quantum circuits:
\begin{equation}
\label{eqtfieldevocirc}
\hat{U}_{1}(\delta t; h_t) =
\begin{aligned}
\Qcircuit @C=-1.3em @R=-0.5cm @! {
& \gate{R_z^{h_t}(\delta t)} & \qw\\
& \gate{R_z^{h_t}(\delta t)} & \qw\\
& \gate{R_z^{h_t}(\delta t)} & \qw\\
& \gate{R_z^{h_t}(\delta t)} & \qw\\
}
\end{aligned}
\end{equation}
and
\begin{equation}
\label{eqhoppingevocirc}
\hat{U}_2(\delta t; J) =  \begin{gathered}\Qcircuit @C=-1.3em @R=-0.5cm @! {
& \ctrl{1} & \gate{R_x^J(\delta t)} & \ctrl{1} & \qw & \qw & \qw & \qw \\
& \targ & \qw                    & \targ & \ctrl{1} & \gate{R_x^J(\delta t)} & \ctrl{1} & \qw \\
& \ctrl{1} & \gate{R_x^J(\delta t)} & \ctrl{1} & \targ & \qw & \targ & \qw \\
& \targ & \qw                    & \targ & \qw & \qw & \qw & \qw \\
}
\end{gathered},
\end{equation}
where $R_x^J(\delta t) = e^{i J \delta t \hat{\sigma}^x}$ and $R_z^{h_t}(\delta t) = e^{i \delta t h_t \hat{\sigma}^z}$. We selected the qubits 0, 1, 2, 3 on Boeblingen so that we could observe the occupation index output over time using a consistent set of qubits for all measurements.

Following the methodology used in~\cite{GustafsonIsing}, the quantum circuit describing the model is initialized
with states that can be interpreted as one or two particle states
and allowed to evolve over a fixed number of Trotter time steps.  The qubit states are recorded at the end of the Trotter time steps and interpreted as accurate expressions for the evolution of the approximate particle occupations.

The choice of parameters $h_T=1$ and $J=0.02$ provides a large gap between the vacuum and the one-particle states which have small energy splittings corresponding to the kinetic energy.  For comparisons and benchmarking purposes, we measure the output of the qubits in the $\sigma^z$ basis, and perform all comparisons using the operator
\begin{equation}
    \hat{n} = (1 - \hat{\sigma}^z) / 2.
\end{equation}
From the point of view of the spectrum, $J$ is a perturbation, but for the real-time evolution of the $\langle n_j(t)\rangle$ with initial states which are eigenstates of $n_j$, this quantity remains constant
in the limit $J=0$. Consequently, the changes in $\langle n_j(t)\rangle$ are driven by $J$.

The choice of the Trotter step depends on the physics goals.  With an ideal quantum computer, algorithmic mitigation can be accomplished by  considering a collection of small time steps controlled by rigorous Trotter bounds such as  a $(\delta t)^2$ proportionality.   However, our long-term objective is to use NISQ machines to calculate the real-time evolution for scattering processes involving few particles.  Small time steps often prevent us from reaching a time scale relevant for the scattering process.

In the specific example considered here, it takes a time of order $t\sim 100$, or $Jt\sim 2$, to go from an initial state $\ket{1000}$ to a state
roughly resembling $\ket{0001}$. On the other hand, the size of the Trotter step necessary to control the Trotter error with rigorous bounds is $\delta t \lesssim 1$. This would require about 100 Trotter steps which is beyond what current quantum computers can achieve.

In the following, we explore the possibility of taking $\delta t$ 10-20 times larger than suggested by a rigorous Trotter bound.  This offers the possibility of reaching relevant time scales, defined as the time needed to bring a free particle to a situation where it interacts with another particle or an external potential.  Having made such a selection it is critical to fully understand what will be the consequences on the output data that such a larger $\delta t$ will have as compared to the smaller time steps suggested to maintain rigorous Trotter bounds.

Following the discussion of the Trotter approximations in  Ref. \cite{treview,ybook},
we examine in detail the one step discrepancy for the simplest Trotter approximation where the discrepancy is of order $(\delta t)^2$
\beq
\Delta_2U \equiv
\e^{-i(h_T \hat{H}_T+J\hat{H}_{JJ})\delta t}-e^{-ih_T \hat{H}_T\delta t}e^{-iJ\hat{H}_{JJ}\delta t}\simeq\frac{h_TJ}{2}[\hat{H}_T,\hat{H}_{JJ}](\delta t)^2.\enq
The bound
\beq
||\Delta_2 U ||\leq \frac{h_TJ}{2}||\ [\hat{H}_T,\hat{H}_{JJ}]\ ||(\delta t)^2,\enq
is sharp for  $\delta t$ small enough.
Numerically we find that for $N_s=4$.
\beq
||\ [\hat{H}_T,\hat{H}_{JJ}]\ ||\simeq 8.944,  \enq
and that the bound is sharp for $\delta t\leq 0.3$.

\begin{figure}[h]
 \centering
\includegraphics[width=0.8\textwidth]{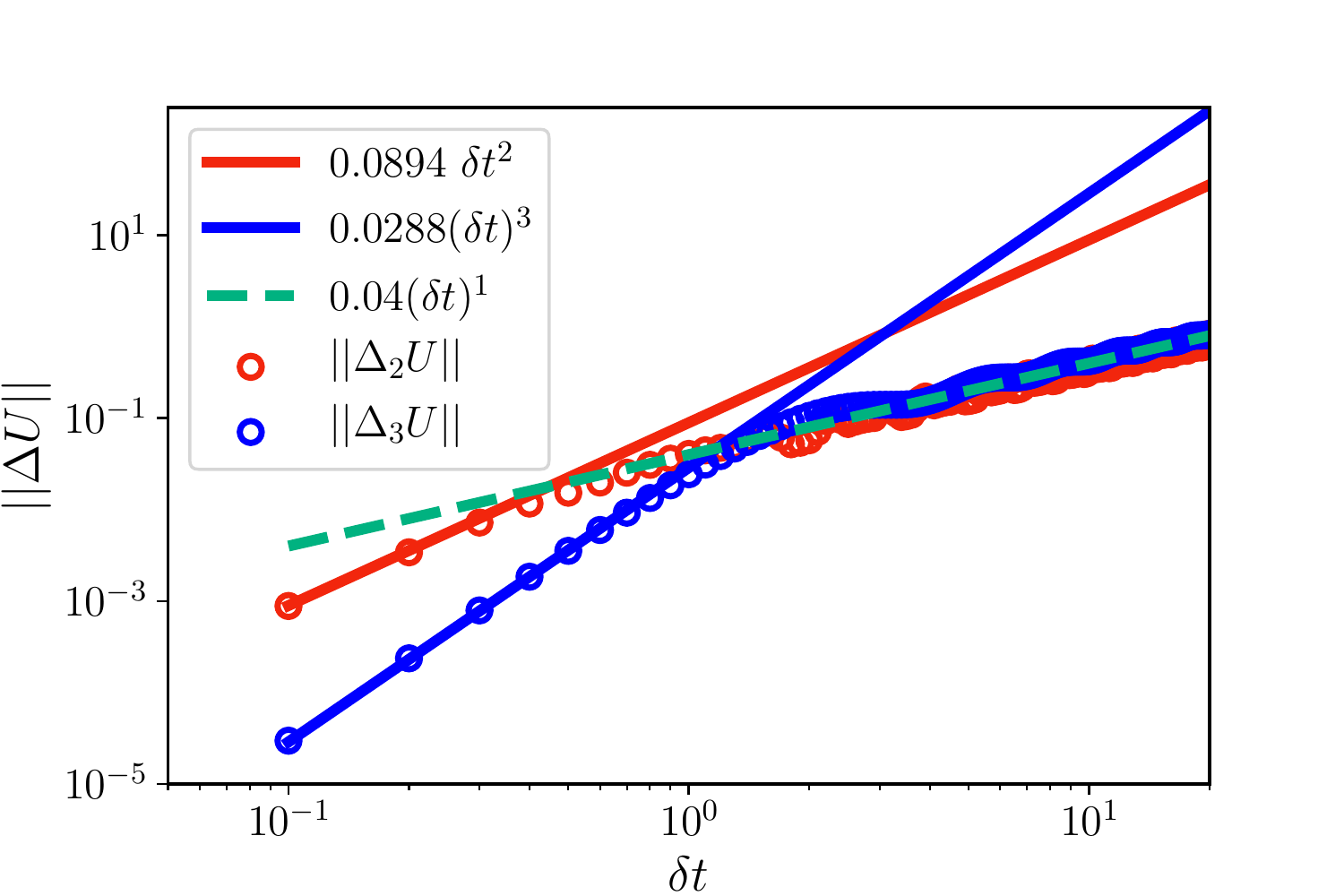}
\caption{$||\Delta_2 U ||$ and $||\Delta_3U ||$versus $\delta t$ for $h_T=1$ and $J=0.02$. }
\label{fig:improvedtrotter}
\end{figure}

For the more accurate Trotter approximation where we have
\begin{eqnarray}
\Delta_3U&\equiv&e^{-i(h_T \hat{H}_T+J\hat{H}_{JJ})\delta t}-e^{-ih_T \hat{H}_T\delta t/2}e^{-iJ\hat{H}_{JJ}\delta t}e^{-ih_T \hat{H}_T\delta t/2} \\&\simeq& \Big(\frac{-h_T^2J}{24}[\hat{H}_T,[\hat{H}_T,\hat{H}_{JJ}]]+\frac{h_TJ^2}{12}[\hat{H}_{JJ},[\hat{H}_T,\hat{H}_{JJ}]]\Big)(\delta t)^3.\end{eqnarray}
We now consider the bound
\beq
||\Delta_3U||\leq \Big(\frac{h_T^2J}{24}||\ [\hat{H}_T,[\hat{H}_T,\hat{H}_{JJ}]]\ ||+\frac{h_TJ^2}{12}||\ [\hat{H}_{JJ},[\hat{H}_T,\hat{H}_{JJ}]]\ ||\Big)(\delta t)^3.\enq
Numerically,
\beq
||\ [\hat{H}_T,[\hat{H}_T,\hat{H}_{JJ}]]\ ||\simeq 35.777,\enq
and
\beq
||\ [\hat{H}_{NN},[\hat{H}_T,\hat{H}_{JJ}]]\ ||\simeq 28.844.
\enq
The bound is sharp for $\delta t\leq 0.7$.
These observations for $||\Delta_2U ||$ and $||\Delta_3U||$ are illustrated in Fig.~\ref{fig:improvedtrotter}. We see that  $\delta t$ small enough we have
\beq
||\Delta_2U ||\equiv
||e^{-i(h_T \hat{H}_T+J\hat{H}_{JJ})\delta t}-e^{-ih_T \hat{H}_T\delta t}e^{-iJ\hat{H}_{JJ}\delta t}||\simeq0.09(\delta t)^2,\enq
and
\beq
||\Delta_3U||\equiv ||e^{-i(h_T \hat{H}_T+J\hat{H}_{JJ})\delta t}-e^{-ih_T \hat{H}_T\delta t/2}e^{-iJ\hat{H}_{JJ}\delta t}e^{-ih_T \hat{H}_T\delta t/2} ||
\simeq 0.03 (\delta t)^3.\enq
However, for For $\delta t \gtrsim$1, we found the empirical bound
\beq||\Delta_2U || \sim ||\Delta_3U ||\sim 0.04 (\delta t)^1.\enq

A detailed numerical analysis shows that Trotter steps of order $\delta t \sim 10$ actually have a reasonable error for ten steps. With current machines, good control can be achieved with 6 Trotter steps
having $\delta t=5$, see Fig. \ref{fig:fulldt5correct} in \ref{sec:machinespecs}.

We collected most of the data for this ``safe" choice of parameters and also explored both increased number of Trotter steps and larger size Trotter steps.  For these choices, the dependence  of observables on $\delta t$ is highly nonlinear and involves resonance phenomena that were not anticipated initially.  As we will explain, these nonlinear effects make algorithmic mitigation problematic for our range of parameters. However, the Trotter approximations that we use are reasonably accurate and we will focus on the closeness of machine measurements to the Trotter approximation implemented on the machines.

Given the size of the lattice, the Hilbert space is sufficiently small ($2^4$ states), that the problem can be completely run on a conventional computer. Using these two calculations for comparison, we call these very accurate results ``Trotter exact" as opposed
to the noisy results obtained with the quantum computers. We now discuss how to quantify the closeness between the two data sets.

\subsection{Benchmarking Measure}
In order to adequately compare results from different hardware platforms, it is crucial to define a metric that globally reflects the closeness of a data set to the numerically accurate Trotter values of the number occupations $\langle n_i(t)\rangle$ at successive times. We found that standard metrics  such as $\chi^2$ or mean squared errors were not completely adequate. The lack of reliable estimates of systematic errors, which are expected to be much larger than statistical errors,  prevents a meaningful use of the $\chi^2$ comparison. Averages of mean-square errors tend to overemphasize the largest errors and do not provide good discrimination among platforms.
We also noticed that the closeness to the numerical values of the Totter values  for qubits with larger values of $\langle n_i(t)\rangle$ were a good indicator for
the quality of the overall evolution, while the qubits with of $\langle n_i(t)\rangle \leq 0.2$ did not play a substantial role.

For these reasons we decided to introduce a new index, that we call the ``$G$-index", that captures  the gross averaged discrepancy.  This index takes all the absolute values of the differences of measured data points from the accurate Trotter  values for all the data points whose measured value is above a certain threshold $\epsilon$ and then averages them:
\begin{equation}
    \label{eq:gad}
    G_\epsilon(data \ set)\equiv \frac{\sum_{n_{meas}>\epsilon}| n_{meas}-n_{Trotter}|}{\sum_{n_{meas}>\epsilon} 1 }.
\end{equation}
As we will see, choices of $\epsilon\simeq 0.2$ keeps the data that contributes to a reasonable discrimination among the different hardware platforms.

\section{Error Mitigation}
\label{secErrorMitigation}
\subsection{Readout Correction}

The magnitude of readout errors (misidentifying a $|1\rangle$ for a $|0\rangle$ or vice-versa) for current superconducting qubit quantum computers can be range above several percent. For this reason, it is important to identify the best method for correcting these readout errors. We examine three well used methods of readout mitigation that are available: two variations of operator rescaling and a calibration matrix method. 

The operator rescaling methods work by using the documented readout errors of a given machine to correct observables via post processing.
This method has two varieties which we will call asymmetric and symmetric. The asymmetric method assumes that there may be some asymmetry in the readout errors, while the symmetric method assumes that the readout error probabilities are symmetric. Both of these methods have a drawback. It is difficult to correct for correlated readout errors because the current formulations below, Eqs. (\ref{eqrobustreadoutfix}) and (\ref{eqnaivereadoutfix}), are not feasibly scalable to large numbers of qubits. The asymmetric readout error mitigation scheme for the Pauli $\hat{Z}$ operator proposed in Ref. \cite{Kandala:2017aa} which accounts for the asymmetric readout errors, is 
\begin{equation}
\label{eqrobustreadoutfix}
\langle \hat{\sigma}^z \rangle = \frac{\langle \hat{ \sigma}^z_{\text{noisy}}\rangle + p_{0\rightarrow1} - p_{1\rightarrow0}}{(1 - p_{0\rightarrow1} - p_{1\rightarrow0})},
\end{equation}
where $\langle \hat{\sigma}^z_{noisy} \rangle$ is the expectation value of the operator as measured on the machine before post processing, $p_{0\rightarrow1}$ is the probability of misidentifying a $|0\rangle$ as a $|1\rangle$ and $p_{1\rightarrow0}$ is the probability of misidentifying a $|1\rangle$ as a $|0\rangle$. This scheme can be approximated assuming that the readout errors are identical, which simplifies the formula in Eq. (\ref{eqrobustreadoutfix}):
\begin{equation}
\label{eqnaivereadoutfix}
\langle \hat{\sigma}^z \rangle \simeq \frac{\langle \hat{\sigma}^z_{\text{noisy}} \rangle}{1 - 2p_{\text{readout error}}}.
\end{equation}

The calibration method implemented here can be found in IBM's published Qiskit library \cite{Qiskit}. The method of correction involves using asymmetric readout errors for the qubits of the quantum computer to construct a readout noise model. Then for each possible state for the system, the readouts are measured and used to construct a probability matrix $\mathcal{M}$ that describes the expected values given some true state. The matrix is then inverted to reverse the probability distribution while subjecting the end result to the constraint that the observables maintain physical values.

                  \begin{table*}
\centering
                  \begin{tabular}{|c|c|c|c|c|}
                  \hline
                   machine & raw & symmetric & asymmetric & calibration \\\hline 

                  Almaden & 86.0(8.3) & 76.4(8.6) & 74(10) & 74(10) \\\hline 
Boeblingen & 67.2(8.0) & 50.8(6.7) & 36.0(6.8) & 31.7(7.2) \\\hline 
Melbourne & 120(15) & 108(15) & 96(11) & 96(11) \\\hline 
\end{tabular}\caption{$G_{\epsilon}$ index $(G_{0.0} \times 10^3)$ summarized for various machines over $\delta t = 5$.}\label{tab:tol0readoutdt5summary}\end{table*} 
                
 \begin{table*}
\centering
                  \begin{tabular}{|c|c|c|c|c|}
                  \hline
                   machine & raw & symmetric & asymmetric & calibration \\\hline 

                  Almaden & 88(11) & 80(11) & 56(16) & 55(16) \\\hline 
Boeblingen & 79(16) & 55.2(8.8) & 16.6(3.0) & 11.2(3.4) \\\hline 
Melbourne & 183(23) & 177(24) & 141(18) & 142(18) \\\hline 
\end{tabular}\caption{$G_{\epsilon}$ index $(G_{0.2} \times 10^3)$ summarized for various machines over $\delta t$ = 5.}\label{tab:tol2readoutdt5summary}\end{table*} 

                  \begin{table*}
\centering
                  \begin{tabular}{|c|c|c|c|c|}
                  \hline
                   machine & raw & symmetric & asymmetric & calibration \\\hline 

                  Almaden & 73.3(8.2) & 64.3(7.3) & 31.2(5.3) &
                  30.6(5.1) \\\hline 
Boeblingen & 101(12) & 64.8(8.2) & 18.4(3.7) & 11.3(4.5) \\\hline 
Melbourne & 216(15) & 209(17) & 158(20) & 159(20) \\\hline 
\end{tabular}\caption{$G_{\epsilon}$ index $(G_{0.3} \times 10^3)$ summarized for various machines over $\delta t$ = 5.}\label{tab:tol3readoutdt5summary}\end{table*}

\begin{figure}[h]
\includegraphics[width=0.48\textwidth]{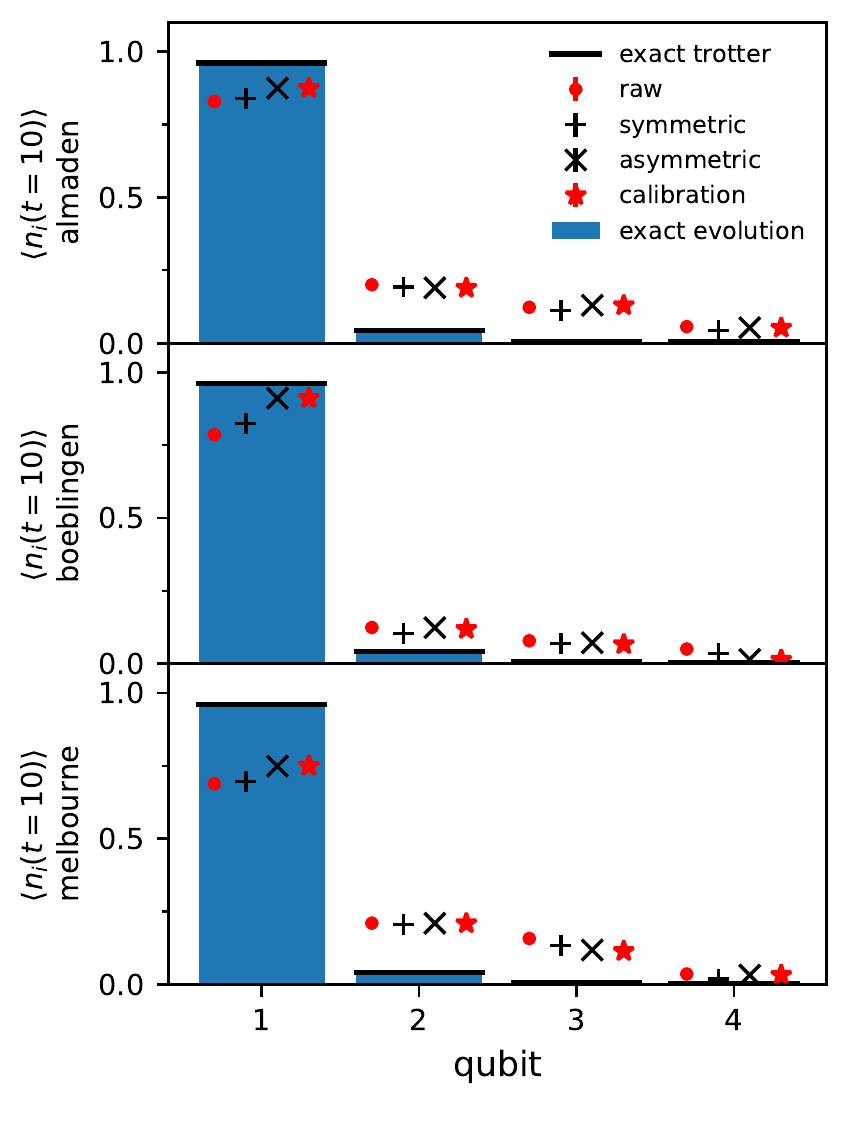}
\caption{Comparison of various readout correction methods (No correction, symmetric correction, asymmetric correction, and calibration matrix) on all machines for $\delta t = 5$ and $t = 10$. The statistical errors are too small to be seen on the figure.}
\label{figReadoutCorrection}
\end{figure}

A comparison of how these different methods work is demonstrated in Fig. \ref{figReadoutCorrection}.
\raggedbottom
\footnote{A list of the asymmetric readout errors can be found in Table \ref{tab:readouterrors} in  \ref{sec:machinespecs}. A complete figure of all $\delta t=5$ times steps can be seen in Fig. \ref{fig:fulldt5correct}.} Readout correction methods, unsurprisingly, improve the accuracy of the results when  using the $G_{\epsilon}$ index. The methods of readout correction from most accurate to least were: the calibration matrix, asymmetric correction, symmetric correction, and then no correction. In particular, Boeblingen has the most significant improvement from  readout correction while the improvements from Almaden and Melbourne were less significant but still noticeable. These results are supported by the metric defined in Tables \ref{tab:tol0readoutdt5summary}, \ref{tab:tol2readoutdt5summary}, and \ref{tab:tol3readoutdt5summary}. Overall the calibration method appears to be the most effective method and we will focus the continuing analysis using just this method. It should also be noted that if we simply take the square root of the mean squared error, all the values are of the order $0.2$. This does not provide sufficient discrimination among the machines and methods.

A sharpening of the differences between these results can be seen by increasing $\epsilon$ in $G_{\epsilon}$ for $\epsilon = 0$ in Table \ref{tab:tol0readoutdt5summary}, $\epsilon = 0.2$ in Table \ref{tab:tol2readoutdt5summary}, and $\epsilon = 0.3$ in Table \ref{tab:tol3readoutdt5summary}.
The increasing of $\epsilon$ provides a better contrast. It both allows comparisons among different machines and helps distinguish the accuracy among different methods on the same machine.
For example in the calibration correction method we can see that by increasing $\epsilon$ the metric lowers for Almaden and Boeblingen while it increases for the Melbourne which has a larger value for $G_{\epsilon}$. This is the desired result of increasing $\epsilon$; the different metric values begin separating so that the differences are clearly discernible. We have examined larger choices of $\delta t$ ($20,~10$, and $20 / 3$), and have included a discussion in \ref{sec:ancillary}. 

\subsection{Gate Noise Mitigation}
In order to minimize the errors introduced by noisy quantum gates, the primary method that is commonly used is a Richardson extrapolation \cite{Dumitrescu:2018njn,Klco:2018zqz,2018arXiv180803623E,PhysRevX.8.031027,PhysRevX.7.021050}.  This method was originally proposed in Ref.~\cite{Richardson} and involves increasing the noise in the system by fixed amounts and then extrapolating to the vertical axis-intercept corresponding to a noiseless value. This process is simple in the case of CNOTs, where an odd number of CNOT gates are inserted into the circuit to increase the overall circuit noise. A simple noise model which admits a linear regression for a noisy expectation value as a function of the noise parameter $\epsilon$ can be derived by noting that the CNOT maps the Pauli group onto itself \cite{Dumitrescu:2018njn}. However, that this does not necessarily eliminate all the systematic noise.

Several different methods of carrying out this extrapolation exist. One method involves polynomial error model fitting as used in \cite{Dumitrescu:2018njn,Klco:2018zqz}. The other leading method, deferred Richardson extrapolation, involves solving a set of $n$ linear equations for $n$ unknown values corresponding to the coefficients of a polynomial, to determine the ``noiseless limit," by way of the Vandermonde matrix \cite{press2007numerical}, as used in \cite{Kandala_2019,2018arXiv180803623E,PhysRevX.8.031027,PhysRevX.7.021050}.\footnote{\cite{PhysRevX.8.031027} also proposed an exponential ansatz but we have not carried out this method due to computational constraints.} Mathematically this works by solving the matrix equation:
\begin{equation}
\label{eq:deferredlimit}
    \mathcal{R}_{i, j} \vec{c}_j = \vec{\mathcal{O}}_i,
\end{equation}
for $\vec{c}$, where $\mathcal{R}_{i, j} = r_i^j$, $r_i$ is the $i^{th}$ error rate $r$, $j$ corresponds to the order of the polynomial  coefficient ranging from $0$ to $n_{points} - 1$, $\vec{c}$ is the coefficient vector for the polynomial terms, and $\vec{\mathcal{O}}$ is a vector of the observable at various error rates. 
The error rates are 1, 3, 5, etc. which correspond to increasing the original noise.

We examine both of these methods as a way of comparison. For fitting, we tested two different ans{\"a}tze to fit the noisy data: a quadratic ansatz in $r$,
\begin{equation}
\label{eqpolyansatz}
\langle \mathcal{O}(t;r)\rangle = A + B r + C r^2,
\end{equation}
and linear ansatz in $r$,
\begin{equation}
\langle \mathcal{O}(t;r)\rangle = A + B r.
\end{equation}
For the deferred limit approach, we use between 2 to 4 different noise rates to carry out this extrapolation. 
\begin{figure}[!ht]
    \centering
    \includegraphics[width=0.5\textwidth]{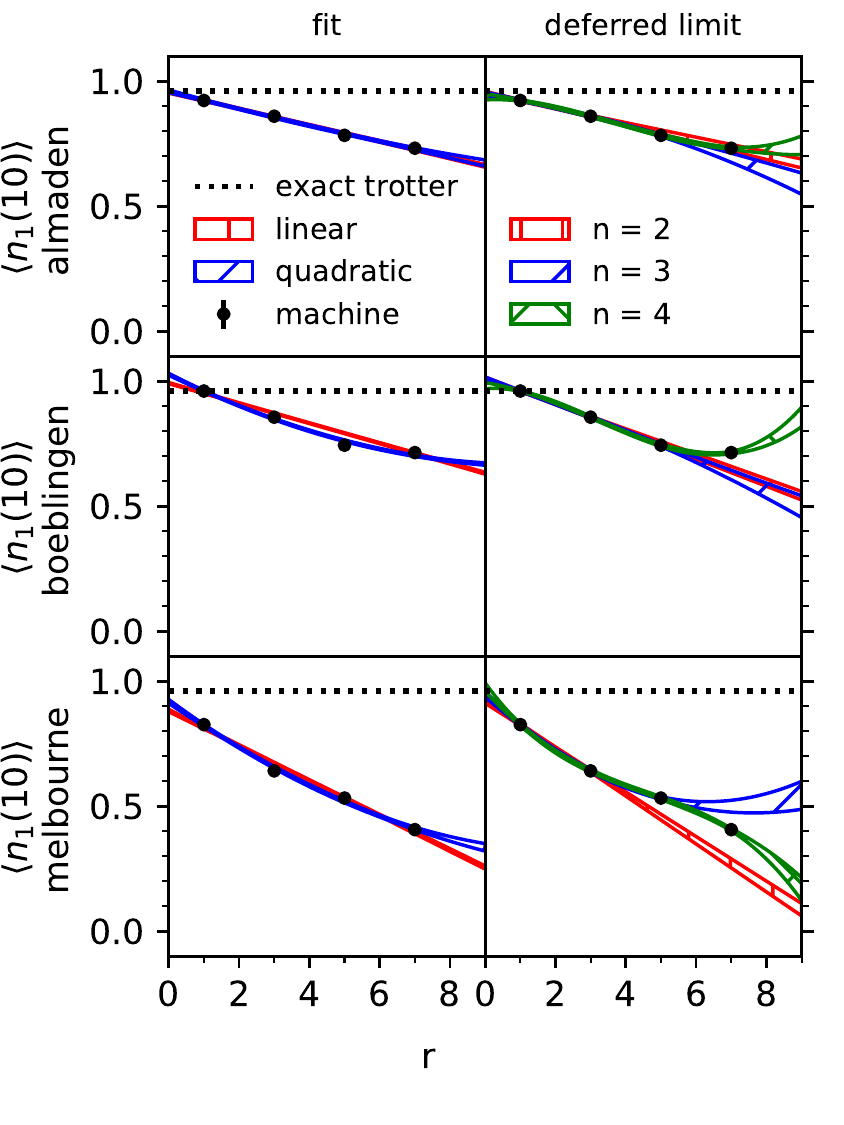}
    \caption{Richardson extrapolations with $\delta t = 5$ for $\langle \hat{n}_1\rangle$ for all three machines using increasing multiples of CNOT gates, corresponding to error rates $r=1,~3,~5,~\text{and }7$.}
    \label{fig:almaden20richardson}
\end{figure}

\begin{figure}
    \centering
    \includegraphics[width=\textwidth]{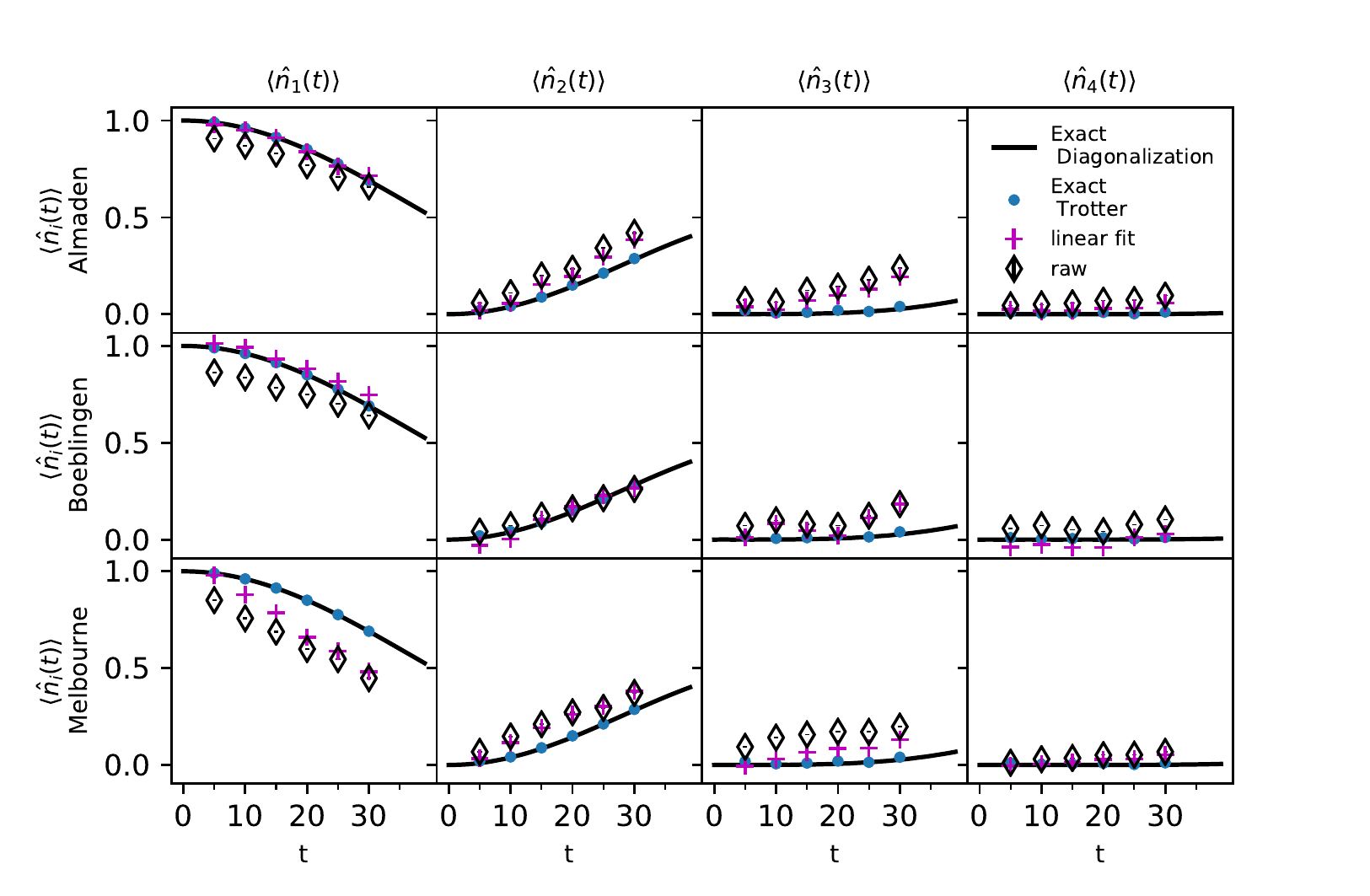}
    \caption{Comparison of machine results ($\delta t = 5$) with (1) no mitigation (raw) $\Diamond$, and (2) calibration readout correction with linear fit noise mitigation (+) to the exact Trotter $\bullet$ and exact continuous evolution (solid line).}
    \label{fig:trottercompare}
\end{figure}

A selected example of these Richardson extrapolations using Almaden is shown in Fig. \ref{fig:almaden20richardson}. There are several main features that are clear from these methods. First, when observables are near the maximally uncertain value, the noise mitigation methods can become somewhat unstable. Second, after several Trotter steps significant errors begin to accumulate and destroy much of the structure of the noise. The issues with error accumulation are demonstrated in Fig. \ref{fig:richardsonsupplementary} in \ref{sec:ancillary}. Finally, the higher order approximations are drastically affected by overfitting. From our results we used a linear approximation for the Richardson extrapolation in the algorithmic mitigation as it appears to be the most accurate. A visual summary of the results for $\delta t = 5$ can be seen in Fig. \ref{fig:trottercompare} and the $G_{\epsilon=0}$ metric is listed for the machines in Table  \ref{tab:richardsonmaindt5}.\footnote{In \ref{sec:ancillary}, Fig. \ref{fig:richardsonsupplementary} shows a case where Richardson extrapolations may not be very effective.}  In two of the three cases (Almaden and Melbourne), the linear fit Richardson extrapolation introduces a noticeable improvement when compared to the raw measurements. One interesting result that is noticeable on the new machines is that overfitting of the noise by using higher order extrapolation is possible and does diminish the accuracy of the extrapolations. 

\begin{figure}
    \centering
    \includegraphics[width=\textwidth]{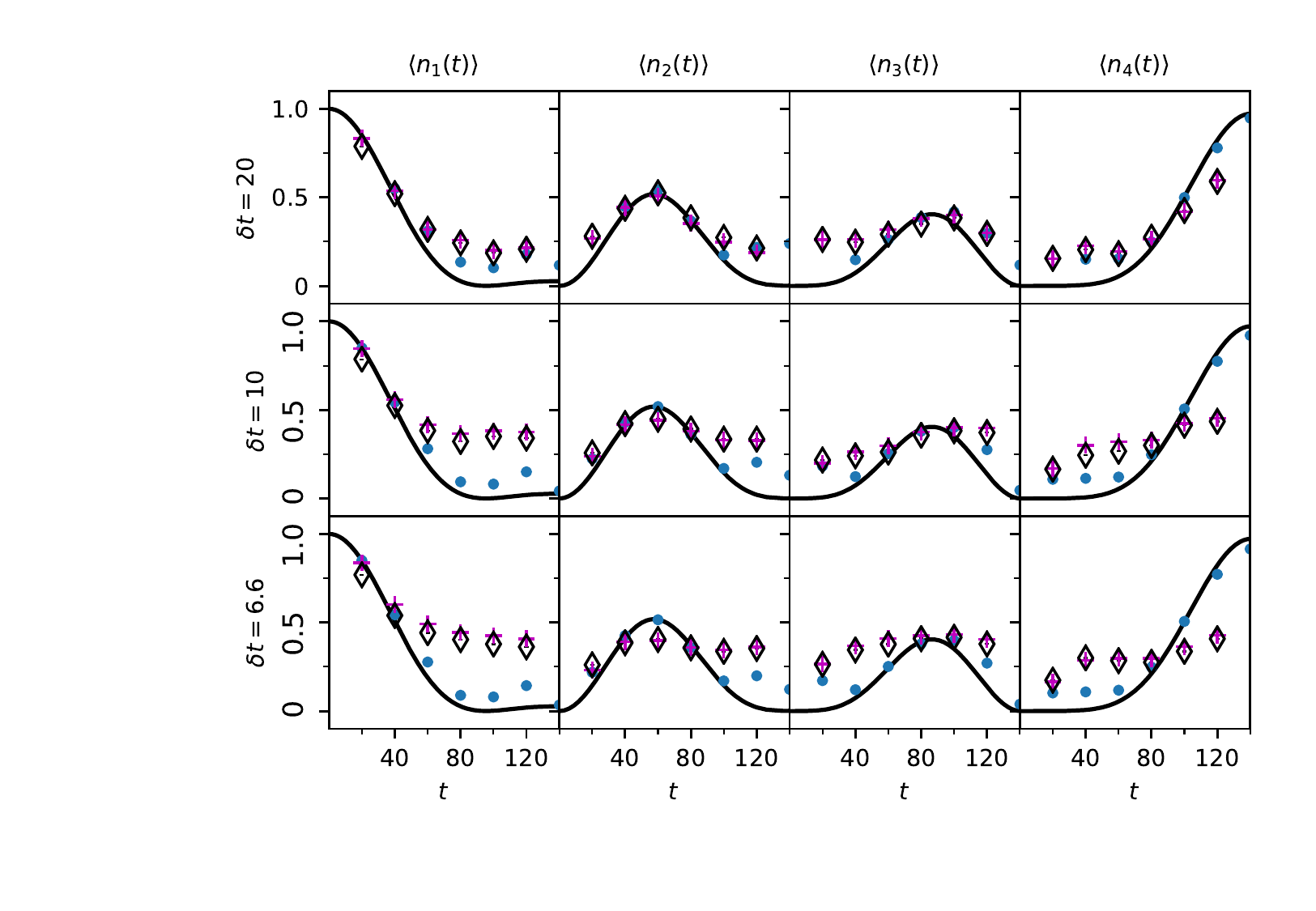}
    \caption{Comparison of raw and Richardson extrapolation using a linear fit for $\delta t = 20$, $10$, and $20 / 3$ on Almaden with (1) no mitigation (raw) $\Diamond$, and (2) calibration readout correction with linear fit noise mitigation (+) to the exact Trotter $\bullet$ and exact continuous evolution (solid line). It is important to keep in mind that for $\delta t = 10$ and $6.6$ are 12 and 18 Trotter steps respectively involved in the full calculation.}
    \label{fig:otherdtrichardson}
\end{figure}

In addition we also ran simulations at longer time frames ($t = 20, 40, 60, 80, 100, $ and $120$) for each of the $\delta t = 20$, $10$, and $20 / 3$. The data from these simulations can be found in Table \ref{tab:richardsonother} and Fig. \ref{fig:otherdtrichardson}. One of the key features we can see is the loss of signal at larger number of Trotter steps, in particular for $\delta t = 10$ and $20 / 3$.

\begin{table*}[!ht]
\centering
\begin{tabular}{|c|c|c|c|c|c|c|c|c|}
\hline
machine & $\delta t$ & raw & calibration & $n = 2$ & $n = 3$ & $n = 4$ & linear & quadratic\\\hline
Almaden & $5$ &76.9(9.2)&65(21)&49(27) &63(33) &81(41) &35(18) &49(26) \\ \hline
Boeblingen &$5$ &57.8(9.0)&25(14)&66(16) &83(29) &107(38) &38(15) &81(19) \\\hline
Melbourne &
$5$ &111(16)&87(25)&41(16) &29(12) &32(12) &69(27) &36(14) \\ \hline
\end{tabular}
\caption{$G_{\epsilon}$ index $(G_0 \times 10^3)$ after Richardson extrapolation for $\delta t = 5$. $n_{points} =2$ (linear) ,$3$ (quadratic), $4$ (cubic) correspond to the various orders in the deferred limit approach defined in Eq. (\ref{eq:deferredlimit}) and linear and quadratic models for the fitting method. $n_{points} = 0$ and $1$ do not have any significant meaning because they correspond to no extrapolations at all.} 
\label{tab:richardsonmaindt5}
\end{table*}
\begin{table*}
\begin{tabular}{|c|c|c|c|c|c|c|c|c|}\hline
machine & $\delta t$ & raw & calibration & $n = 2$ & $n = 3$ & $n = 4$ & linear & quadratic\\\hline
Almaden & $20$ &44.4(7.8)&39(14)&58(24) &87(37) &120(48) &40(16) &63(25) \\ \hline
Boeblingen & $20$ &89(15)&88(32)&127(38) &163(61) &208(86) &116(40) &133(41) \\ \hline
Melbourne & $20$ &155(24)&164(53)&181(61) &181(64) &177(66) &169(58) &193(60) \\ \hline \hline
Almaden & $10$ &138(21)&141(43)&107(40) &93(37) &105(31) &149(50) &104(40) \\ \hline
Boeblingen & 
$10$ &173(32)&171(69)&160(87) &179(86) &209(85) &184(79) &182(86) \\ \hline
Melbourne &
$10$ &172(32)&174(65)&197(78) &223(85) &243(89) &168(71) &205(80) \\ \hline\hline
Almaden & $20/3$ &174(23)&181(47)&160(49) &150(45) &142(44) &184(55) &159(48) \\ \hline
Boeblingen &
$20/3$ &230(37)&249(82)&257(102) &286(102) &321(101) &249(91) &259(108) \\ \hline
Melbourne &
$20/3$ &169(33)&170(66)&177(76) &182(78) &176(79) &167(75) &189(77) \\ \hline
\end{tabular}
\caption{$G_{\epsilon}$ index $(G_0 \times 10^3)$ for Almaden after Richardson extrapolation for $\delta t = 20,~ 10,$ and $20 / 3$.}
\label{tab:richardsonother}
\end{table*}

When analyzed together, the data in the tables and figures referenced in this section point toward a physical picture of what is happening. From Fig. \ref{fig:trottercompare} and the $G_0$ metric listed for the machines in Table  \ref{tab:richardsonmaindt5} we can see that for $\delta t = 5$ there is a clear improvement in the fidelity of the output data between the raw data and linear fit.  In addition the occupation number results for each of the qubits track the expected Trotter theoretical evolution for each of the IBM Q hardware platforms on which this code was run.

Increasing the value of $\delta t = 20$, $10$, and $20 / 3$, as shown in Table \ref{tab:richardsonother} indicates that the improvements between the raw and linear fits have disappeared on each of the IBM Q hardware platforms.  Essentially what is happening is that the linear and quadratic error models are not capable of fully modelling the noise in the circuits.  This is also confirmed in Fig. \ref{fig:otherdtrichardson} plots illustrating the deviation of the measured data (taken from Almaden) from the theoretical expectation value of the Trotterization for each of the occupation numbers.  

Finally, it is noted that these results do not track to what would naively be expected from referencing the quantum volume \cite{PhysRevA.100.032328} associated with each machine.  The problem is that the quantum volume is a metric for quantum computer performance that is intended to measure the size of a quantum processor's accessible state space.  Given the fact that a processor with $n$ qubits has a $2^{n}$ dimensional state space the maximum theoretical number of computational states that can be accessed is $2^{n}$.  In reality, that number is considerably less.  The quantum volume metric measures that fraction of the accessible state space by constructing random circuits.  A processor's quantum volume is defined by the ability of a quantum computer to reliably run a family of square random circuits with a width based on the number of qubits and a depth based on the number of steps in that circuit. 

This type of square circuit is not characteristic of most quantum algorithms.  Most algorithms actually have specific shapes (in terms of width and depth) that may vary widely depending on the application being run.  Therefore, quantum volume based on these square random circuit constructs is not necessarily an optimal measure of a quantum computer's performance for a specific application.  This effect has also been recently noticed by several other groups \cite{BlumeKohout2019AVF},  \cite{Resch2019BenchmarkingQC}.  A brief discussion of machine specifications, performance, and quantum volume can be found in \ref{sec:machinespecs}.

\section{Algorithmic Mitigation}
\label{secAlgorithmicMitigation}

After the Richardson extrapolation, we expect that the errors on observables depend only on the size of the Trotter step $\delta t$. For $\delta t$ small enough, the errors after a certain number of Trotter steps are controlled by $\delta t$. However, for the larger values of $\delta t$ considered in Sec. \ref{subsec:trotter}, the situation is more complicated. 

\begin{figure}
  \begin{subfigure}[b]{0.5\textwidth}
    \includegraphics[width=\textwidth]{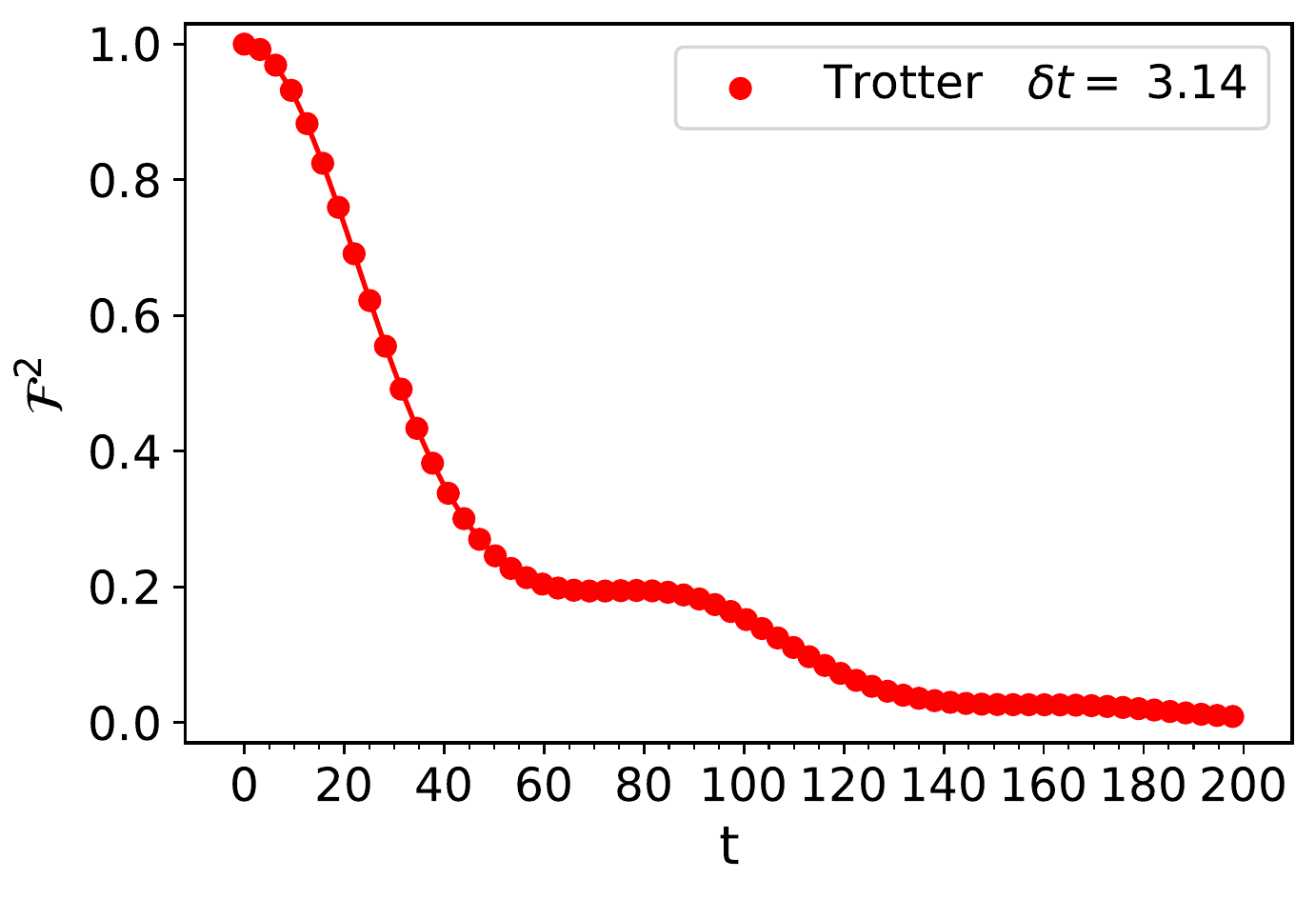}
    \caption{$\mathcal{F} ^2$ versus t for Trotter  $\delta t$ = 3.14 }
    \label{ndeltat3p14}
  \end{subfigure}
  \begin{subfigure}[b]{0.5\textwidth}
    \includegraphics[width=\textwidth]{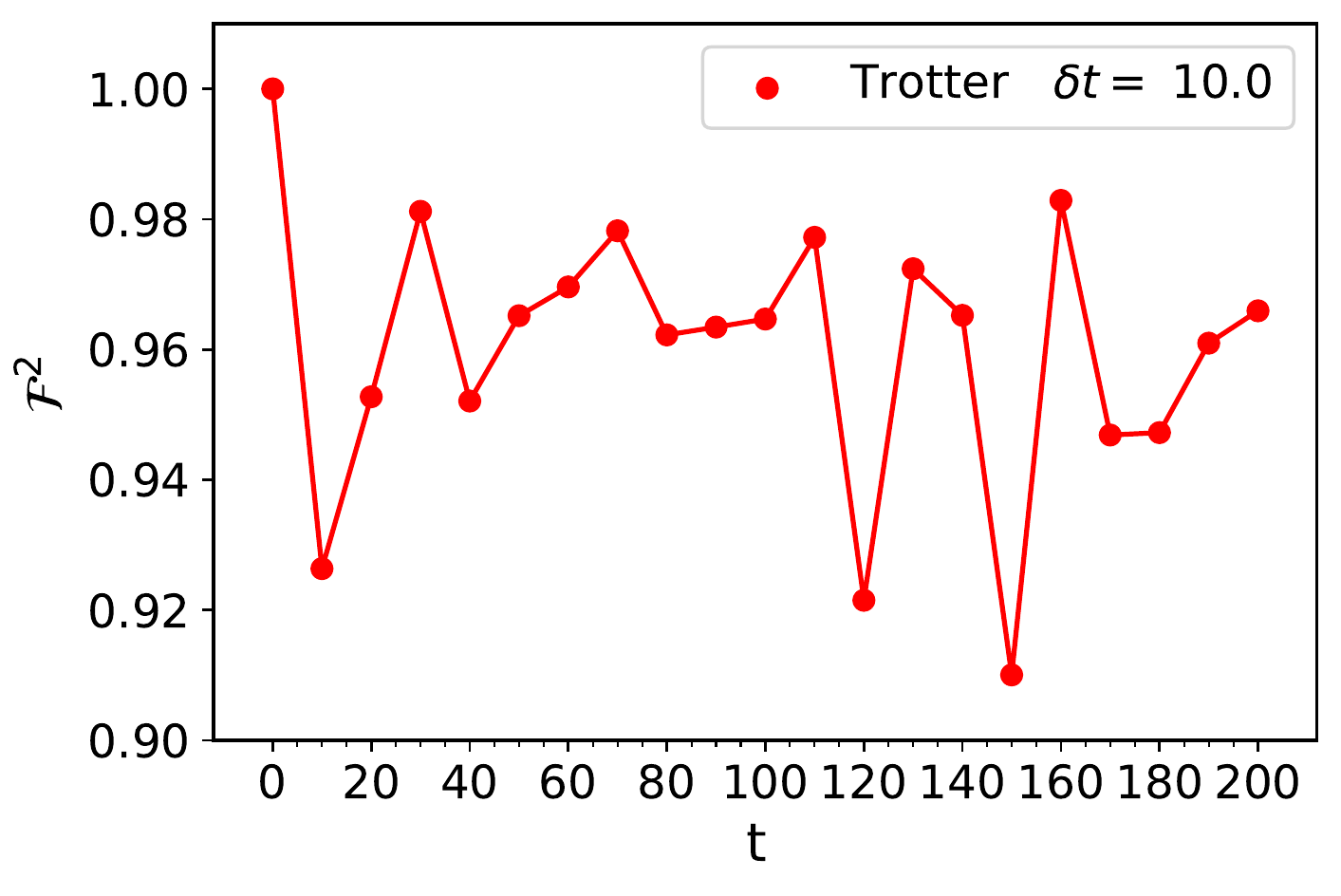}
    \caption{$\mathcal{F} ^2$ versus t for Trotter  $\delta t$ = 10.0 }
    \label{ndeltat10}
  \end{subfigure}
  \caption{Comparison of fidelities for $\delta t$ near-resonant (a) and away from resonance (b).}
\end{figure}

\begin{figure}
    \centering
    \includegraphics[width=0.5\textwidth]{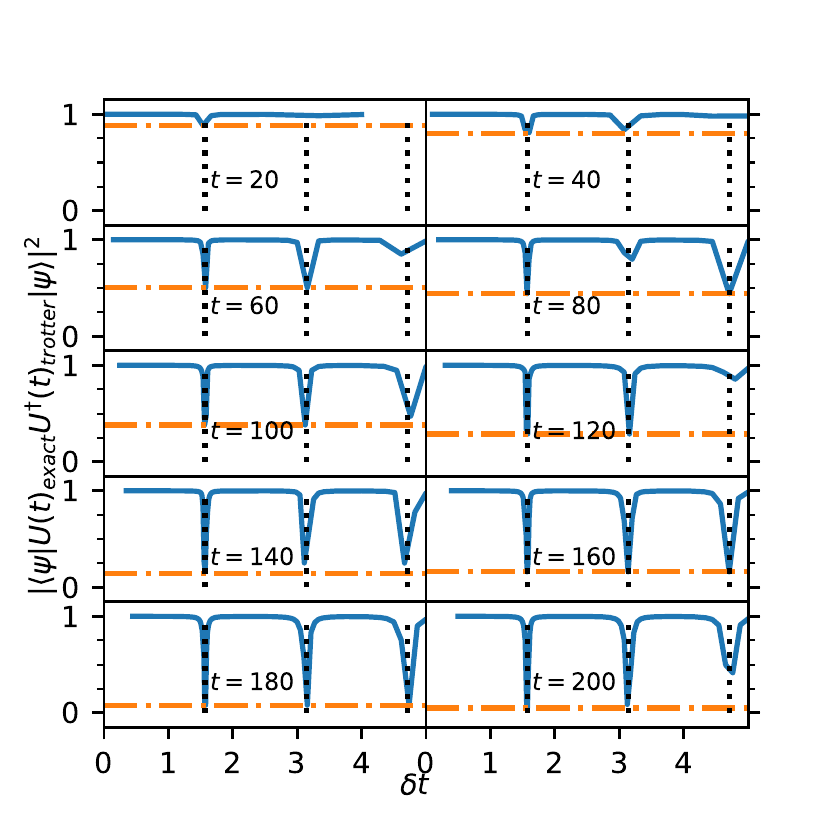}
    \caption{Fidelity, $|\langle \psi|U(t)^{\dagger}_{exact} U(t)_{Trotter}|\psi \rangle|^2$, of the Trotter operator as a function of $\delta t$. Blue line: fidelity; dot dashed orange line: peak minimum; dashed black line: center of peak.}
    \label{fig:fidelitytrotter}
\end{figure}

Certain choices of $\delta t$ can cause a sudden loss of accuracy in the Trotterization.  Evidence of these ``resonant" values of $\delta t$ can be seen by calculating the square of the fidelity,
\begin{equation}
\label{eq:fidelity}
\mathcal{F}^2\equiv|\langle1000| U(t)_{exact}^\dagger U_{Trotter}(t)|1000\rangle|^2,
\end{equation} 
versus $t$ and  comparing  Fig. \ref{ndeltat3p14} for $\delta t$ near a resonance at 3.14 versus Fig. \ref{ndeltat10} with a $\delta t$ of $10$.
This fidelity is affected because near-resonant choices of $\delta t$ will cause one or more terms in the Trotterization to be close to the identity. In addition, shrinking the Trotter step increases the number of Trotter steps to reach a given $t$ and correspondingly increases the actual machine errors. For $\delta t < \pi / 2$, the Trotterization is tightly constrained by the rigorous bounds and free of resonance. It is then safe to assume a ``Richardson" extrapolation as used in Sec. \ref{secErrorMitigation} \cite{PhysRevA.99.012334}. This can be explained by the fact that for $h_T$ = 1, the spectrum of the second term of the Hamiltonian has  eigenvalues which are even integers and consequently $\hat{U}_2(\pi/2;1)$ is minus the identity and the second term actually disappears from the Trotter approximation of the evolution of the observable. As shown in Sec. \ref{subsec:trotter}, for $\delta t \gtrsim 1$, the one-step Trotter error grows much slower than the rigorous bounds, however the lack of smoothness near integer multiple of $\pi/2$ makes mitigation difficult (see Fig. \ref{fig:fidelitytrotter}). 
This Richardson extrapolation can be carried out by using either polynomial fits or the Richardson's deferred limit approach, where $r$ now scales like $(\delta t) / (\delta t)_{smallest}$. However because of these ``resonant" points, an algorithmic extrapolation will encounter trouble as the effects of larger $\delta t$'s will be problematic, which had not been appreciated at the onset of this work.

While in theory evolving to the same time with a small Trotter step size addresses this issue, the inherent noise from the quantum gates prevents this result from being achieved in actuality. Methods for working around and understanding these effects so that algorithmic extrapolations can be used will be a focus of future work when sufficiently small $\delta t$ can be achieved.

\section{Summary and Next Steps}
\label{secConclusion}

The results reported here demonstrate that it is possible for the time evolution of the transverse Ising model with four sites to be accurately simulated for limited time scales on current generation NISQ based superconducting transmon machines. In addition they also show that applying basic methods for machine noise reduction can improve the examination of time evolution beyond one or two Trotter steps.  These successes are tempered by the well recognized facts that basic quantum error mitigation techniques typically can only provide limited error mitigation for a few species of well-known errors.  Several of these sources of errors are discussed here 

The T1 and T2 values and the single and two qubit error rates listed in the hardware platform's backend properties only provide a partial insight into the full spectrum of errors found on NISQ systems.  As mentioned in the text, the systematic errors on these systems are difficult to measure with the tools that have been applied to the problem up to this point.  These sources of error originate through the impacts of spectator qubits that contribute to the total error when they are activated by microwave pulses directed to other qubits in the system.  This degrades the accuracy of the measured qubit results through coherent errors and cross talk.  These type of errors may be as large as the simple error measurements from the one qubit and two-qubit error rates themselves.  Simple one and two qubit calibrations of the qubits directly involved in the circuit will not properly capture these errors.  It also was recognized at the outset that there may also be intra-day and inter-day re-calibration drift of the qubit properties and that these instabilities may impact the occupation index measurements and the subsequent data analysis.  

The results of the Richardson extrapolation applied to the physics models running on today's quantum computers are undercut by the previously mentioned difficulties of the algorithmic error mitigation strategies. 

It is clear that there is some efficiency in the algorithmic mitigation and that its success is dependent upon slight improvements in two qubit gate fidelity and judicious choices of Trotter steps. Nevertheless, a proper control of algorithmic errors would require significantly smaller Trotter steps than the ones used here.  However, in order to perform a Richardson extrapolation more precisely, it will require implementing error mitigation strategies at the hardware pulse level.  Follow-on work is underway in this area.

The issue of increasing the value of $J$ should not be noticeably more difficult than the current simulations. The Trotter step time $\delta t$ will naturally need to be reduced because the Trotter error scales as a product of $J \delta t$ as is discussed in Sec. \ref{secTheory}. The second consideration is that at larger $J$, ``pair creation" effects will become significant and therefore the ability to disentangle these different ``particle sectors" will be important.

Despite these recognized error mitigation issues, the results reported here indicate progress addressing these problems.  This methodology is the first to apply the combined set of quantum computing techniques and error mitigation to real-time measurements for basic physics models running on NISQ based machines.  It has provided valuable insights toward our understanding how to model these types of physics problems on superconducting transmon quantum computing hardware platforms. This methodology can naturally be extended to the Thirring and Schwinger models as both of these models have Hamiltonians which can be written in terms of local tensor products of Pauli-matrices \cite{lamm2019parton,Muschik:2016tws,Klco:2018kyo}. Specifically the tensor products are of $\hat{\sigma}^x\hat{\sigma}^x$, $\hat{\sigma}^y\hat{\sigma}^y$, and $\hat{\sigma}^z\hat{\sigma}^z$ which can be easily implemented using current quantum computing technology.  The results demonstrated here with current generation NISQ based machines indicate that these methodologies have the potential to be extended to calculations of phase shifts in basic physics models.  We are now planning future work that includes a more sophisticated in-depth error mitigation study using cycle benchmarking, randomized compiling, and the $G_{\epsilon}$ to illustrate the impacts of systematic and coherent errors on the transverse Ising model and other field theory Hamiltonians.  

\section{Acknowledgments}
This work was supported in part by the U.S. Department of Energy (DoE) under Award Number DE-SC0019139. Dreher was supported in part by the U.S. Department of Energy (DoE) under award DE-AC05-00OR22725. We thank Raphael Pooser for a careful reading of the manuscript and many helpful comments. We thank North Carolina State University (NCSU) for access to the IBM Q Network quantum computing hardware platforms through the NCSU IBM Q Hub.

\bibliography{IOPmain.bbl}

\appendix

\section{G-index analysis for larger Trotter steps}
\label{sec:ancillary}
In order to examine the physics of real-time scattering modelled in this 1+1 field theory on these IBM Q platforms, larger Trotter steps must be used in order to access the region where the computation represents a more physically interesting time evolution. For this reason, simulations at $\delta t = 20,~10,$ and $20 / 3$ were used to study the time evolution beyond the initial $\delta t = 5$ results averaged over the three $\delta t$.  For each of these calculations, the G-Index was computed.  Tables \ref{tab:tol0readoutsummary}, \ref{tab:tol2readoutsummary}, and \ref{tab:tol3readoutsummary} illustrate the results for $\epsilon = 0.0$, $\epsilon = 0.2$, and $\epsilon = 0.3$ respectively.  

A key feature that emerged from these computations shows that increasing $\epsilon$, sharpens the discrimination among the machines in terms of the noise mitigation.  From among all of the data analyzed, Almaden show the best improvement on a relative basis at the value of $\epsilon = 0.3$.  A second key feature shows that results of longer time scale evolution continue to indicate that the applied error mitigation methods will not effectively address the errors within the current readout error mitigation methods.  Nevertheless, the data do allow some statements to be made as to a clear delineation in machine efficacy.

Additional noise mitigation analysis was performed on the output data from the larger Trotter steps using the Richardson correction procedure.  Table \ref{tab:richardsonother} shows the results from these runs at each of the larger $\delta t$ values with both linear and quadratic algorithmic fits.  What emerges from these results is that the Richardson extrapolation method is ineffective at these larger Trotter steps for all of the machines on which the field theory model was run.  In some cases, these algorithmic mitigation procedures actually deteriorate the final results, indicating that the non-linearity in the noise is not well modelled by simplified linear or quadratic algorithmic error mitigation.

\begin{table*}
\centering
                  \begin{tabular}{|c|c|c|c|c|}
                  \hline
                   machine & raw & symmetric & asymmetric & calibration \\\hline 

                  Almaden & 91.2(8.0) & 86.3(8.0) & 89.5(8.4) & 89.4(8.4) \\\hline 
Boeblingen & 124(13) & 119(13) & 116(14) & 118(15) \\\hline 
Melbourne & 152(13) & 151(13) & 150(13) & 150(13) \\\hline 
\end{tabular}\caption{$G_{0.0}\times 10^3$ summarized for various machines over $\delta t$ = 20, 10, 20/3.}\label{tab:tol0readoutsummary}

                  \begin{tabular}{|c|c|c|c|c|}
                  \hline
                   machine & raw & symmetric & asymmetric & calibration \\\hline 

                  Almaden & 69(10) & 66(10) & 61.9(9.9) & 61.8(9.9) \\\hline 
Boeblingen & 102(14) & 99(15) & 92(15) & 91(16) \\\hline 
Melbourne & 157(19) & 159(20) & 152(19) & 152(19) \\\hline 
\end{tabular}\caption{$G_{0.2}\times10^3$ summarized for various machines over $\delta t$ = 20, 10, 20/3.}\label{tab:tol2readoutsummary}
                  \begin{tabular}{|c|c|c|c|c|}
                  \hline
                   machine & raw & symmetric & asymmetric & calibration \\\hline 

                  Almaden & 71(14) & 69(14) & 55(14) & 54(14) \\\hline 
Boeblingen & 116(20) & 110(21) & 99(22) & 94(22) \\\hline 
Melbourne & 190(26) & 193(26) & 177(25) & 178(25) \\\hline 
\end{tabular}\caption{$G_{0.3}\times10^3$ summarized for various machines over $\delta t$ = 20, 10, 20/3.}\label{tab:tol3readoutsummary}\end{table*}

\begin{figure*}
    \centering
    \includegraphics[width=\textwidth]{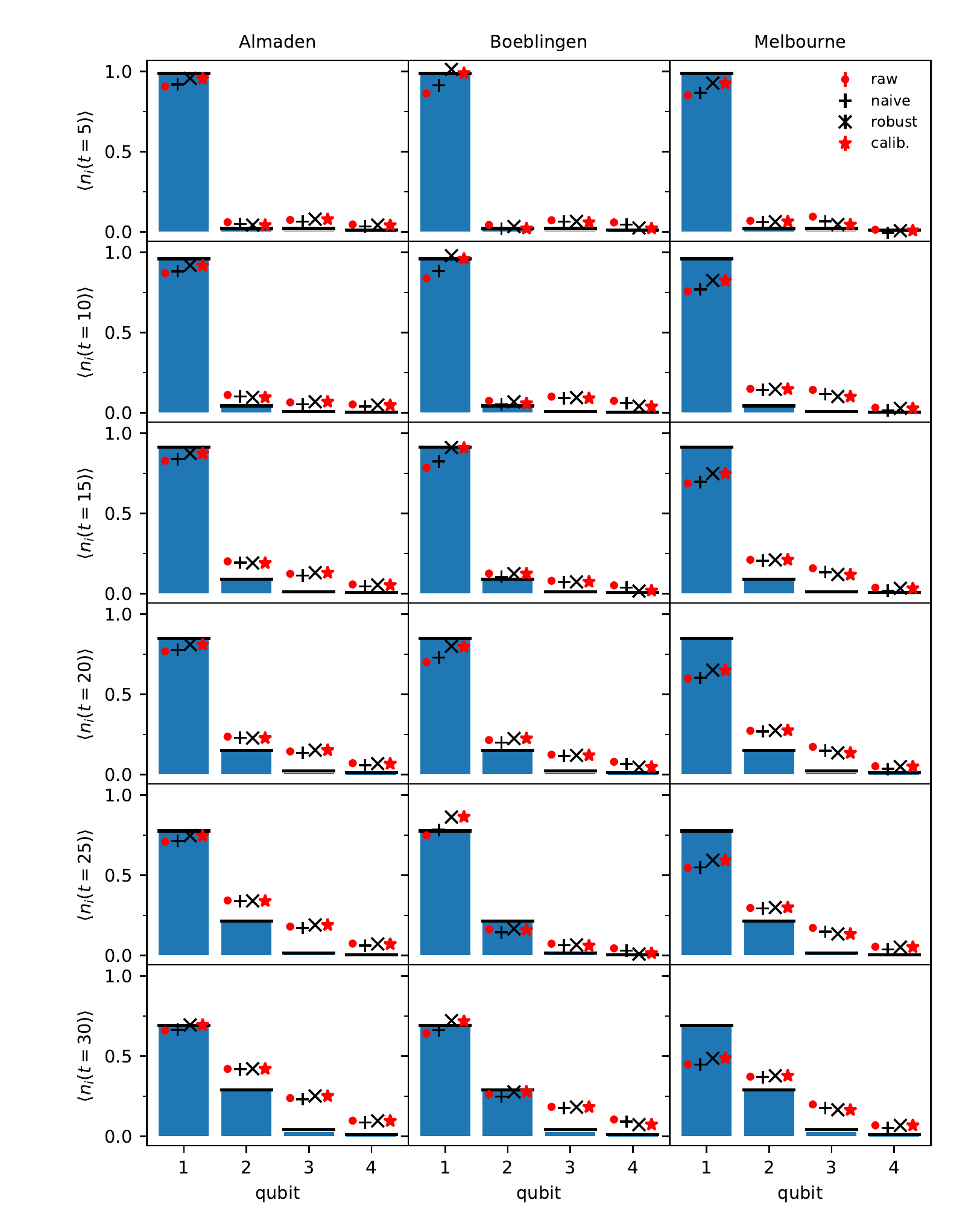}
    \caption{Comparison of various readout correction methods across all three machines for $\delta t = 5$.}
    \label{fig:fulldt5correct}
\end{figure*}
\begin{figure}
\includegraphics[width=0.5\textwidth]{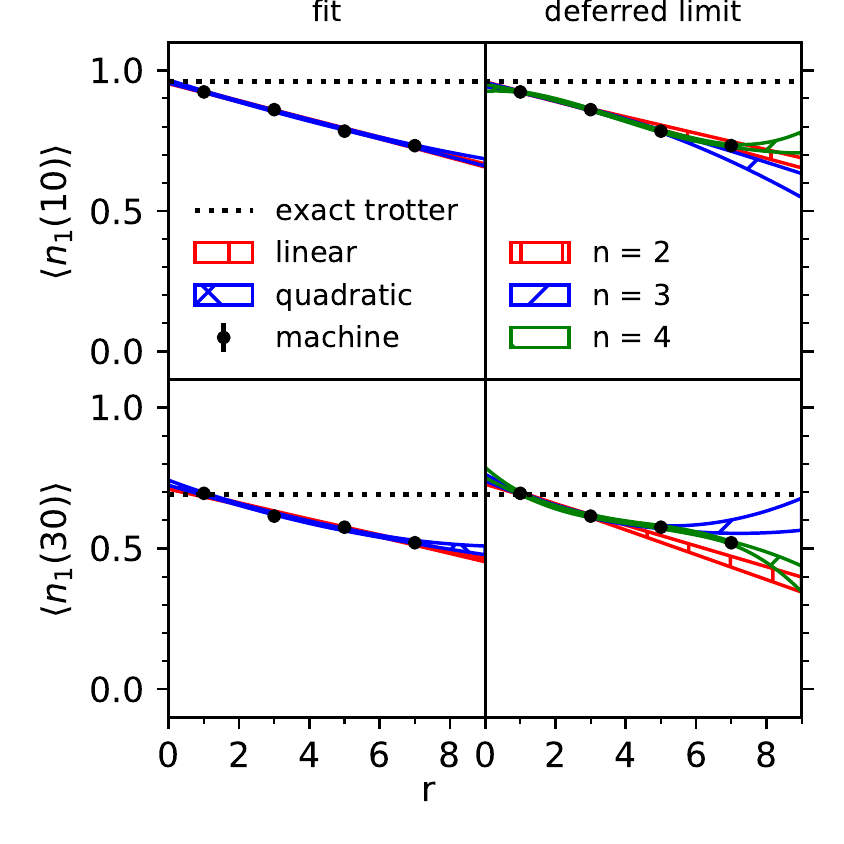}
\caption{Richardson Extrapolations for two different time steps ($t = 10$ and $t = 30$) using Almaden with $\delta t = 5$.}
\label{fig:richardsonsupplementary}
\end{figure}

\section{Machine Specifications and Performance}
\label{sec:machinespecs}
\begin{figure}[ht]
    \centering
    \includegraphics[width=0.5\textwidth]{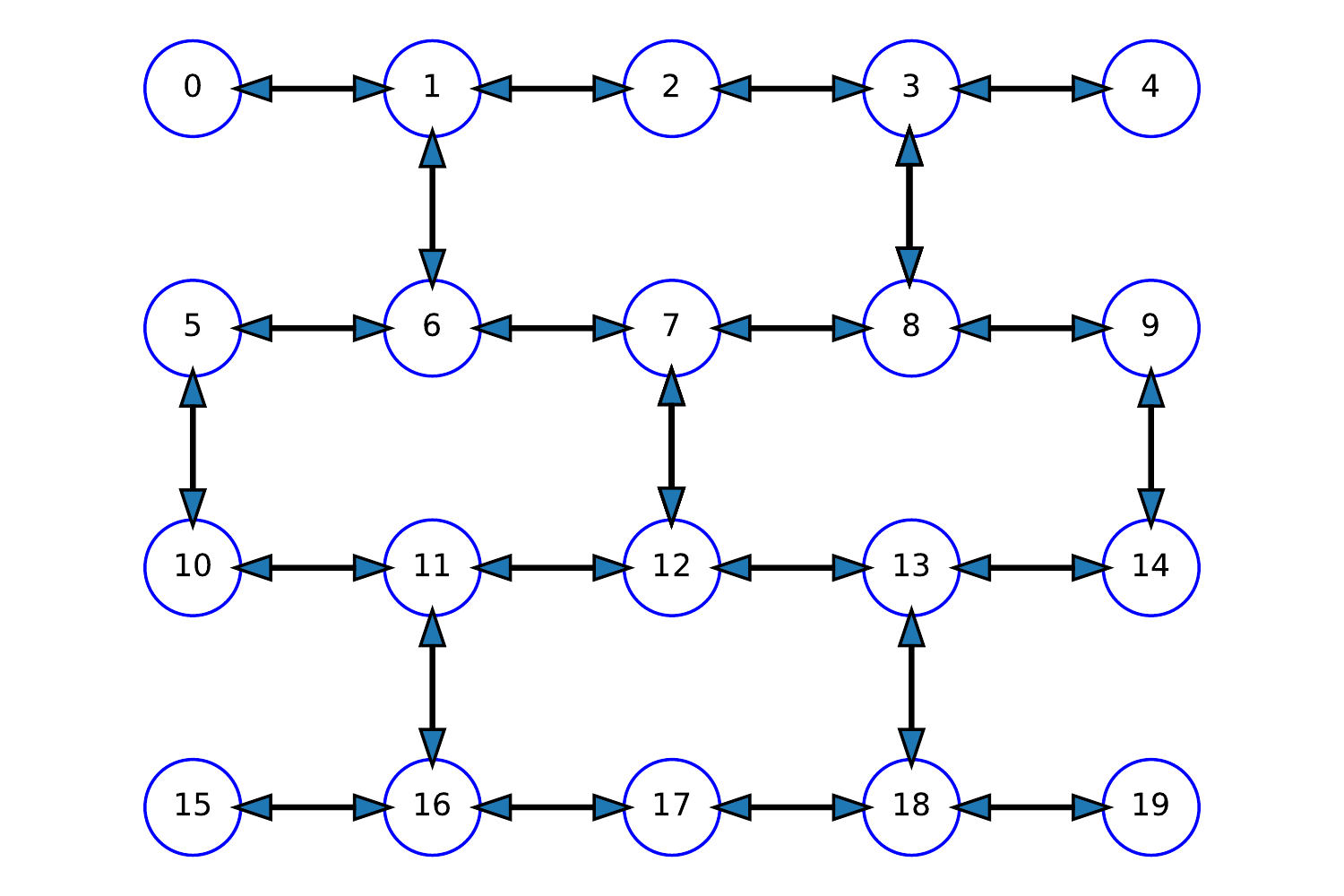}
    \caption{Qubit Layout on Almaden and Boeblingen}
    \label{fig:almadenlayout}
    \includegraphics[width=0.5\textwidth]{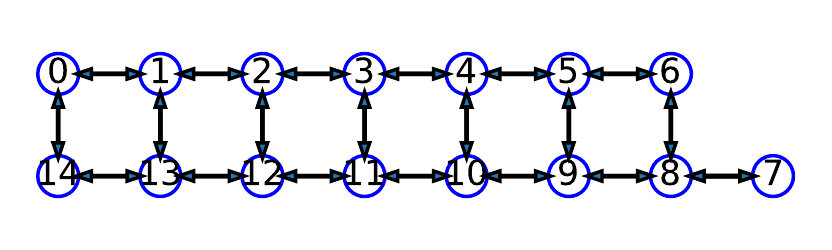}
    \caption{Qubit Layout on Melbourne}
    \label{fig:melbournelayout}
\end{figure}

Three different IBM Q hardware platforms were used for this project.  In this appendix, we provide information regarding the machine specifications and hardware layout. The three machines chosen were Melbourne (14 qubit machine introduced into service in September 2018), Boeblingen (20 qubit machine introduced into service in August 2019), and Almaden (20 qubit machine introduced into service in September 2019).  Figure \ref{fig:almadenlayout} and Figure \ref{fig:melbournelayout} show the layouts for the different machines.  

IBM recently developed a single number metric called the "quantum volume" \cite{PhysRevA.100.032328}. This technique uses randomized model circuits to measure improvement in system-wide gate error rates for near-term quantum computation and error-correction experiments.  This metric is then used as a relative measure when comparing different IBM Q hardware platforms. The published quantum volumes for each machine are as follows; Melbourne has QV = 8, Boeblingen has QV = 16 and Almaden has a QV = 8.  Other information provided by IBM with regard to the machines used in this study is given in Table \ref{tab:readouterrors} which lists the asymmetric readout errors for the machines on which the simulations were run. 

\begin{table*}
    \centering
    \begin{tabular}{|c|c|c|c|c|}\hline 
    qubit & property & Almaden & Boeblingen & Melbourne\\ \hline
   1 &  $P(0\rightarrow1)$ & 0.0067 & 0.0933 & 0.005 \\ \hline
   1 &  $P(1\rightarrow0)$ & 0.0533 & 0.1467 & 0.083 \\ \hline
   2 &  $P(0\rightarrow1)$ & 0.0200 & 0.0133 & 0.008 \\ \hline 
   2 &  $P(1\rightarrow0)$ & 0.0300 & 0.0900 & 0.028 \\ \hline 
   3 &  $P(0\rightarrow1)$ & 0.0000 & 0.0100 & 0.055 \\ \hline 
   3 &  $P(1\rightarrow0)$ & 0.0533 & 0.0333 & 0.078 \\ \hline 
   4 &  $P(0\rightarrow1)$ & 0.0066 & 0.0367 & 0.006 \\ \hline 
   4 &  $P(1\rightarrow0)$ & 0.0500 & 0.0267 & 0.068 \\ \hline 
    \end{tabular}
    \caption{Readout Error probabilities as listed in IBM's machine backend on QISKIT.}
    \label{tab:readouterrors}
\end{table*}

Based on the topology of the qubit layout showing the connectivity of the qubits for both Almaden and Boeblingen does not allow for a trivial implementation of the four site transverse Ising model with periodic boundary conditions such as the model implemented in \cite{GustafsonIsing}. For this reason, open boundary conditions were used instead. For the computations described here a value of $J \delta t = 0.1$ were used because this choice reproduces the exact evolution with a reasonable accuracy for a small set of Trotter steps and is of a sufficient length of time that the real-time dynamics are observable. 

\begin{figure}[!ht]
\centering
\includegraphics[width=0.5\textwidth]{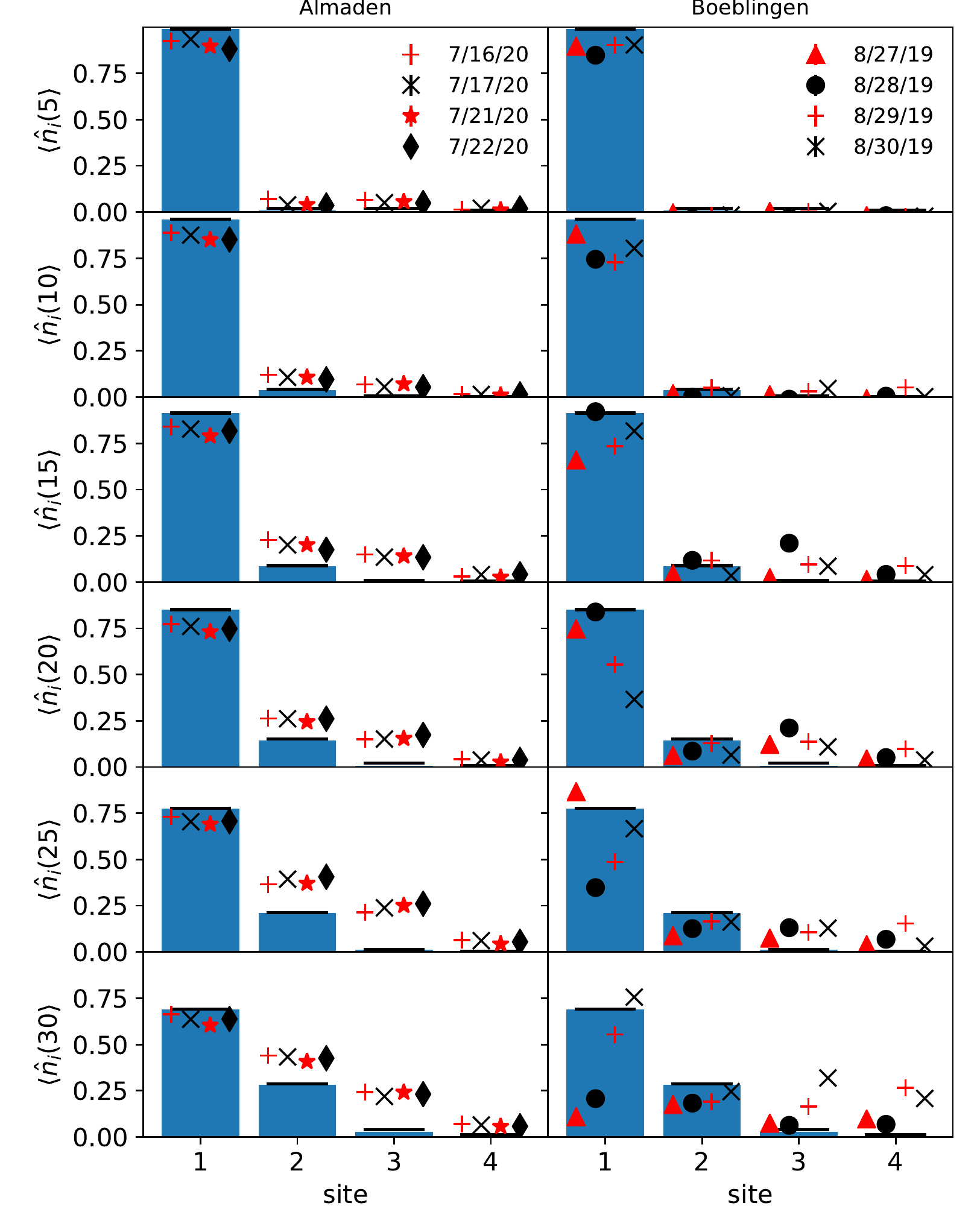}
\caption{\small Comparison of Trotter evolution of the Ising model on the Almaden and Boeblingen machines across multiple days. $J = 0.02$, $h_T = 1.0$, $N_s = 4$, and $\delta t = 5$. The left column corresponds to the Almaden machine; the right column corresponds to the Boeblingen quantum computer.}
\label{figmultipledays}
\end{figure}

Simulations on the Boeblingen and Almaden machines ( Fig. \ref{figmultipledays}) were run across four different dates with fixed parameters to gauge how the performance of the machine changes from day to day. The Boeblingen quantum computer is able to implement between four and five Trotter steps before significant gate errors start to accumulate and distort the wavefunction. It was observed that the Almaden machine is able to implement between four to six of these Trotter steps before gate errors become a noticeable problem.

One particular topic that is worth noting is the relative performance of the three machines is somewhat different from what would be expected based on a quantum volume measurement.   From  quantum volume basis, one would assume that Boeblingen would produce the best results from the $\delta t$ computations.  However, the data in Tables \ref{tab:tol0readoutsummary}, \ref{tab:tol2readoutsummary}, and \ref{tab:tol3readoutsummary} do not support such a conclusion.  This may indicate that the quantum volume procedure based on error rates using randomized model circuits may not be the best metric for characterizing various subject domain applications.  This is currently being investigated in more detail and will be reported in a future publication.

It is worth noting that the errors in the simulations are more driven by gate errors than decoherence errors. This is seen in Table \ref{tab:coherenceerrors} where the chance of an error occurring roughly increases linearly but by 5 to 6 Trotter steps the simulation is already nearing the coherence limit. These errors are on the same order of magnitude for the simulations of interest. 

\begin{table*}
\begin{tabular}{|c|c|c|c|c|}
\hline
steps & CX gates & prob. 1 gate error & average computation time ($\mu s$) & decoherence prob. \\\hline
1&6&0.063&1.844&0.043\\\hline
2&12&0.121&3.688&0.085\\\hline
3&18&0.176&5.532&0.124\\\hline
4&24&0.228&7.376&0.162\\\hline
5&30&0.276&9.22&0.198\\\hline
6&36&0.321&11.064&0.233\\\hline
\end{tabular}\caption{average gate errors, computation times, and decoherence probability, taken on QISKit backend on 7 /10/2020.}
\label{tab:coherenceerrors}
\end{table*}

\end{document}